\documentstyle[12pt]{article}
\begin{document}

\begin{titlepage}
\rightline{September 1993}
\rightline{McGill/93-46}
\vskip 2cm
\centerline{\bf \large
Neutrino oscillations and the exact parity model}
\vskip .8cm
\centerline{R. Foot}
\vskip .7cm
\centerline{{\it Physics Department,
McGill University,
3600 University street,}}
\centerline{{\it
Montreal, Quebec, Canada, H3A 2T8}}
\vskip 2cm

\centerline{Abstract}
\vskip 1cm
\noindent
We re-examine neutrino oscillations in exact parity
models. Previously it was shown in a specific model that
large neutrino mixing angles result. We show here that this is
a general result of neutrino mixing in exact parity
models provided that the neutrino
mass matrix is real. In this case, the effects
of neutrino mixing in exact parity models is such that
the probability of a given weak eigenstate remaining
in that eigenstate averages to {\it less} than a half
when averaged over many oscillations. This result is interesting
in view of the accumulating evidence for a
significant deficit in the number of  solar neutrinos.
It may also be of relevance to the atmospheric neutrino
anomaly.

\end{titlepage}
{\bf \large I Introduction}

\vskip 1cm

\noindent
Neutrino physics is one field of research
which may provide the first indication of physics
beyond the minimal standard model.
Indeed, there are already two exciting indications
that the minimal standard model is incomplete.
Firstly, there are the solar neutrino experiments [1].
For a long time there has been a discrepancy between
the experimental measurements of Davis and
the theoretical predictions. There are now four experiments
which we summarise below:
$$\begin{array}{ll}
Homestake&0.27 \pm 0.04(exp)\pm 0.03(theor),\\
Kamiokande&0.50 \pm 0.05(stat)\pm 0.06(sys)\pm 0.07(theor),\\
Gallex&0.66 \pm 0.11(stat)\pm0.05(sys)\pm 0.03(theor),\\
SAGE&0.44^{+0.13}_{-0.18}(stat)\pm 0.11(sys)\pm 0.02(theor),
\end{array}
\eqno(1)$$
where the data has been normalized to the theoretical prediction
of Bachall et al [2]. Note that the theoretical prediction
involves a lot of assumptions, and the true theoretical
value may be outside these errors. In particular an analysis
by S. Turck-Chieze et al [3] gives theoretical predictions
for the experiements which are quite different to those
of Bahcall but which still seem to be to high to be consistent
with the data (although it has been argued that the data and
the theory may be in-agreement if one takes into account
all sources of uncertainty [4]). In view of the above, we do
not attempt to propose a particular physics solution which will make
all of the experiments agree with the theoretical prediction of
Bahcall et al [2]. If there is a solar neutrino problem and if
new particle physics is the solution then any new particle physics
which can reduce the number of solar neutrinos by a large fraction
(e.g. 1/2) may be the cause of the apparent disagreement
of theory with data.

Another experiment which seems to be inconflict with theory
is the atmospheric neutrino experiment [5]. This experiment measures
the ratio of $\nu_{\mu}/\nu_e$ interactions where the neutrinos
are presumed to originate from cosmic ray
interactions in the atmosphere. These experiments observe a deficit
in the ratio of $\nu_{\mu}/\nu_e$ interactions when the data
is compared with theory. We summarise the situation below:
$$\begin{array}{ll}
Kamiokande&0.60\pm0.07(stat)\pm0.05(sys),\\
IMB&0.55\pm 0.05(stat)\pm 0.10(sys),
\end{array}
\eqno (2)$$
where the data has been normalized to the theoretically expected
ratio [6].

One obvious way to try and reconcile these experiements with
the standard model is to extend the model to incorporate
neutrino masses [7]. Once neutrinos have masses,
then oscillations of neutrino flavours can occur [7].
However, the effect of vaccum oscillations tends
to be small unless there are large mixing angles.
If neutrino masses follow the pattern suggested by the
quarks, then mixing is small, and it is not possible to understand
the solar neutrino (or atmospheric neutrino anomaly) in terms
of vacuum oscillations. Neutrinos coming from the sun travel
through matter and it is possible that the neutrino oscillations are
matter enhanced (MSW effect [8]). This solution to the
solar neutrino problem is probably the most economical solution
in the sense that it proposes the least amount of new
physics. This does not necessarily mean it is the correct
solution (although it may be). Also, one cannot use the MSW effect
to explain the atmospheric neutrino anomaly. If this
anomaly is a signal for new partilce physics, then the most natural
explanation for it is the existence of vacuum neutrino oscillations.
It is possible that vacuum neutrino oscillations is also the
physical mechanism responsible for the solar
neutrino deficit. This is what we assume in this note.
One of the theoretical problems with this interpretation
is that it is hard to understand why the neutrino
oscillations are so big?

Recently, it was pointed out [9] that if the minimal
standard model is extended so that an exact discrete Parity
symmetry exists [9,10,11] (which is unbroken by the vaccuum),
then in that framework it is quite natural for
neutrino flavours to be maximally mixed combinations
of ordinary and mirror matter.
Mirror matter is essentially sterile (when probed
with ordinary matter) so that if the oscillation lengths
are small enough, then the vacuum neutrino oscillations
will reduce the neutrino flux by a factor of two.
In the previous work [9], the considerations
were restricted to a specific model. {\it Here we show that
if the neutrino mass matrix is real, then in general,
vacuum neutrino oscillations will deplete a weak eigenstate by
at least a factor of two.} This is an interesting result
in view of the solar neutrino and atmospheric
neutrino experiments.

The outline of this paper is as follows: In section II we briefly
review the exact parity model. In section III we show that
if neutrinos have mass and the ordinary and mirror neutrinos
mix together, then in general the oscillations will be large.
In fact, we show that the effects of neutrino mixing in exact
parity models is such that the probability of a given weak
eigenstate remains in that eigenstate is less than 1/2 when it
is averaged over many oscillations (provided that the neutrino
mass matrix is real).
In section IV we look at some examples. We examine some
exact parity versions of the see-saw model.
In section V we conclude with some comments.
\vskip 1cm
\noindent
{\large \bf II Exact parity models}

\vskip .7cm
\noindent
In this section we quickly review exact parity symmetric
models [9, 10, 11].
To understand how parity might be conserved, consider
a model which successfully describes present experiments.
In particular, consider the minimal standard model.
This model is described by a Lagrangian ${\cal L}_1$.
This Lagrangian is not invariant under the usual parity
transformation so it seems parity is violated. However,
this Lagrangian may not be complete.
If we add to ${\cal L}_1$ a new Lagrangian ${\cal L}_2$
which is just like ${\cal L}_1$ except that all left-handed
(right-handed) fermions are replaced by new right-handed (
left-handed) fermions which feel new interactions of the same
form and strength, then the theory described by
${\cal L} = {\cal L}_1 + {\cal L}_2$ is invariant under
a parity symmetry (under this symmetry ${\cal L}_1
\leftrightarrow {\cal L}_2$). In addition to these Lagrangian terms,
there may also be terms which mix ordinary with mirror
matter and which are parity invariant.
We label this part of the Lagrangian as ${\cal L}_{int}$.
The terms in ${\cal L}_{int}$ are very important since they lead
to interactions between ordinary and mirror matter, and hence
allow the idea to be experimentally tested in the laboratory.

If we apply the above proceedure to the standard model then
${\cal L}_1$ is just the standard model Lagrangian.
We now add the ``mirror matter'' as described above,
so that the total Lagrangian consists of
two parts ${\cal L}_1$ and ${\cal L}_2$.
Then the gauge symmetry of the theory is
$$SU(3)_1 \otimes SU(2)_1 \otimes U(1)_1 \otimes
SU(3)_2 \otimes SU(2)_2 \otimes U(1)_2. \eqno (3)$$
There are two sets of fermions, the ordinary particles
and their mirror images, which transform under the
gauge group of Eq.(3) as
$$\begin{array}{ll}
f_L \sim (1, 2, -1)(1, 1, 0),& F_R \sim (1, 1, 0)(1, 2, -1),\\
e_R \sim (1, 1, -2)(1, 1, 0),& E_L \sim (1, 1, 0)(1, 1, -2),\\
q_L \sim (3, 2, 1/3)(1, 1, 0),&Q_R \sim (1, 1, 0)(3, 2, 1/3),\\
u_R \sim (3, 1, 4/3)(1, 1, 0),&U_L \sim (1, 1, 0)(3, 1, 4/3),\\
d_R \sim (3, 1, -2/3)(1, 1, 0),&D_L \sim (1, 1, 0)(3, 1, -2/3),
\end{array}
\eqno (4)
$$
(with generation index suppressed). The Lagrangian is invariant
under the discrete $Z_2$ parity symmetry defined by
$$\begin{array}{c}
x \rightarrow -x,\  t \rightarrow t,\\
G_1^{\mu} \leftrightarrow G_{2\mu},\
W_1^{\mu} \leftrightarrow W_{2\mu},\ B_1^{\mu}
\leftrightarrow B_{2\mu},\\
f_L \leftrightarrow \gamma_0 F_R,\ e_R \leftrightarrow
\gamma_0 E_L,\ q_L \leftrightarrow \gamma_0 Q_R,\
u_R \leftrightarrow \gamma_0 U_L,\ d_R \leftrightarrow
\gamma_0 D_L, \end{array}
\eqno (5) $$
where $G_1^{\mu}(G_2^{\mu}), W_1^{\mu} (W_2^{\mu}) $
and $B_1^{\mu} (B_2^{\mu})$ are the gauge bosons of
the $SU(3)_1$ $(SU(3)_2)$,
$SU(2)_1 (SU(2)_2), U(1)_1 (U(1)_2)$
gauge forces respectively.
The minimal model contains two Higgs doublets which are
also parity partners:
$$ \phi_1 \sim (1, 2, 1)(1, 1, 0),\  \phi_2 \sim (1, 1, 0)(1, 2, 1).
\eqno (6)$$

An important feature
which distinguishes this parity conserving theory
from other such theories (e.g. the usual left-right symmetric
model) is that the parity symmetry is assumed to be
unbroken by the vacuum.
The most general renormalizable Higgs potential
can be written in the form
$$V(\phi_1, \phi_2) = \lambda_1 (\phi_1^{\dagger} \phi_1
+ \phi_2^{\dagger} \phi_2 - 2u^2)^2 + \lambda_2 (
\phi_1^{\dagger} \phi_1 - \phi_2^{\dagger} \phi_2)^2, \eqno (7)$$
where $\lambda_{1,2}$ and $u$ are arbitrary constants. In
the region of parameter space where $\lambda_{1,2} > 0$,
$V(\phi_1, \phi_2)$ is non-negative and is minimized
by the vacuum
$$\langle \phi_1 \rangle = \langle \phi_2 \rangle =
\left( \begin{array}{c}
0\\
u
\end{array}\right).
\eqno (8)$$
The vacuum values of both Higgs fields are exactly the same
and hence parity is not broken by the vacuum in this
theory.

If the solar system is dominated by the usual particles,
then the theory agrees with present experiments.
The idea can be tested in the laboratory because it is possible
for the two sectors to interact
with each other via ${\cal L}_{int}$. In the simplest case that we
are considering at the moment (where ${\cal L}_1$ is
the minimal standard
model lagrangian), there are just two possible terms (i.e.
gauge invariant and renormalizable) in ${\cal L}_{int}$.
They are,
\vskip .3cm
\noindent
(1) The Higgs potential terms $\lambda \phi_1^{\dagger} \phi_1
\phi_2^{\dagger} \phi_2$ in Eq.(7) and
\vskip .3cm
\noindent
(2) The gauge boson kinetic mixing term $\alpha
F^1_{\mu \nu} F^{2 \mu \nu}$
where $F^{1,2}_{\mu \nu} = \partial_{\mu} B^{1,2}_{\nu} -
\partial_{\nu} B^{1,2}_{\mu}$.
\vskip .3cm
\noindent
The principle phenomenological effect of the term in (1) is to modify
the interactions of the Higgs boson. This effect will be tested
if or when the Higgs scalar is discovered. The details have been
discussed in Ref.[9]. The principle phenomenological
effect of the kinetic mixing term in (2) is to give small
electric charges to the mirror partners of the ordinary charged
fermions.This effect has also been discussed previously [10,11].

\vskip 1cm
\noindent
{\large \bf III General considerations}
\vskip .7cm
\noindent
In the past there has been much speculation that neutrinos
are massive. If this is the case then this will lead to
another signature of exact parity symmetric gauge models.
This is because ${\cal L}_{int}$ can contain neutrino
mass terms which mix the ordinary and mirror matter. (Note
that if electric charge is conserved,
then it is not possible for ${\cal L}_{int}$ to contain
mass terms mixing the charged fermions of ordinary matter
with mirror matter, however neutrinos may be neutral so
such mass terms are possible providing that the neutrinos
have masses). The actual number of neutral Weyl fermions is not
known, (although the number of weakly interacting
light $SU(2)_L$ doublet neutrinos has been determined to
be $3$ by the LEP experiments) so we will start with
some general results.

Suppose there are $n$ ordinary neutrino Weyl fields,
and $n$ mirror neutrino Weyl fields\footnote{Note that apriori it
is possible to have parity diagonal neutrino fields
(i.e. with the parity transformation $N_R \leftrightarrow
\gamma_0 (N_R)^c$).
If these fields exist then the following arguement is still
valid and Eq.(11) would still be true in this case.} .
Denote these weak eigenstate fields by
$$\nu_{1L}, \nu_{2L}, ...,\nu_{nL},$$
$$N_{1R}, N_{2R}, ..., N_{nR}, \eqno (9)$$
where $\nu_{iL} \leftrightarrow \gamma_0 N_{iR}$ ($i = 1,...,n$)
under the parity symmetry (Eq.(5)).
The physics becomes clearer if we work in the parity diagonal
basis,
$$\nu_{iL}^{\pm} = {\nu_{iL} \pm (N_{iR})^c \over \sqrt{2}}.
\eqno (10)$$
Then, under parity, $\nu_{iL}^{\pm} \rightarrow
\pm \gamma_0 (\nu_{iL}^{\pm})^c$.
Observe that mass terms of the form
$m_{ij} \bar\nu_{iL}^{+} (\nu_{jL}^{+})^c + H.c.$ and
$m_{ij}^{'}\bar\nu_{iL}^{-}(\nu_{jL}^{-})^c + H.c.$
are invariant under parity if $m_{ij}$ and $m_{ij}^{'}$
are real, while terms of the form
$m_{ij}^{''}\bar\nu_{iL}^{+} (\nu_{jL}^{-})^c + H.c.$
are parity odd if $m_{ij}^{''}$ is real,
and hence they are not invariant under the discrete
symmetry (provided $m_{ij}^{''}$ are real). Under this reality
assumption, it follows that the $\nu^{+}$ fields do
not mix with the $\nu^{-}$ fields.
Thus every mass eigenstate neutrino field will
be linear combinations of $\nu^{+}$ fields {\it only},
or linear combinations of $\nu^{-}$ fields {\it only}.
{}From this result one can easily show that an ordinary
weak eigenstate such as the left-handed neutrino field
coupling to the left-handed electron and the $W$ boson
will have the form:
$$\nu_{eL} =  {\nu^{+}_1 + \nu^{-}_1
\over \sqrt{2}},$$
$$      =  \sum_{i=1}^n {\alpha_i \nu_i^{'} \over \sqrt{2}} +
{\beta_i \nu_i^{''} \over \sqrt{2}},
\eqno (11) $$
where
 $\nu_i^{'} (\nu_i^{''})$
are the set of mass eigenstates which are mixtures of the
$\nu_i^{+} (\nu_i^{-})$ fields and
$\alpha_i,$ and $\beta_i,$  are parameters
which depend on the details of the neutrino
mass matrix, but must satisfy $\sum_{i=1}^n
\alpha_i^2 = \sum_{i=1}^n \beta_i^2 = 1  $.

Define ${\cal P}(E,R)$ to be the probability that an initial weak
eigenstate electron neutrino (with energy $E$) remains a weak
eigenstate neutrino after travelling a distance $R$.
Then:
$${\cal P}(E,R) = |\langle \nu_{eL}|\nu_{eL}(E,R)\rangle|^2.
\eqno (12)$$
Define ${\cal P}$ to be the averaged probability:
$${\cal P} = {limit \over R \rightarrow \infty}{1 \over R} \int^R_0
{\cal P}(E, R) dR. \eqno (13)$$
Calculating  ${\cal P}$ we find\footnote{Note that we have assumed
that there are no degeneracies among the masses of the
physical neutrinos.}:
$${\cal P} = \sum^n_{i=1} {\alpha_i^4 \over 4} +
{\beta_i^4 \over 4}.\eqno (14)$$
Hence
$${1 \over 2n} \leq {\cal P} \leq {1 \over 2}. \eqno (15)$$

In the case of the solar neutrino problem, one must average
${\cal P}(E,R)$ by taking into account the region of emission
of the sun, the region of absorption on Earth, and
the energy spectrum of the source. In general such an average will
depend on the mass differences of the physical neutrinos.
However if the oscillation lengths are small enough then the average
probability ${\cal P}$ (defined in Eq.(13)) will be a good
aproximation to the fraction of electron neutrinos reaching
the Earth. In particular, for the case of 2 state mixing,
the averaged
function $\cal P$ is 1/2. In this simple case,
for the neutrino
energies detected by the solar neutrino experiments, the neutrino
flux will be reduced by the factor ${\cal P} = 1/2$
provided that the neutrino mass squared
difference satisfies [2]
$$0.014\ eV^2 > \Delta (m^2) > 3 \times 10^{-10}\ eV^2. \eqno (16)$$
Note that the upper bound ($0.014\ eV^2$) comes from laboratory
experiments [12]. Observe that Eq.(16) represents
a very large range of parameters (8 orders of magnitude).
Thus there is actually a very large range of parameters in which
the flux of electron neutrinos coming from the sun will
be significantly reduced by a factor of 2.
\vskip 1cm
\noindent
{\bf \large IV Some examples: The see saw model.}
\vskip .7cm
\noindent
As an example, consider the exact parity version of
the see-saw model.
The see-saw model is a simple way to understand the
smallness of the masses of the known (i.e. the three left-handed)
neutrinos. We will actually consider two different versions of the
see-saw model to illustrate the results of the last section.
The first version of the see-saw model
that we will use is the one with an additional right-handed
gauge singlet neutrino (together with its mirror image)
per generation. Before we consider the
parity invariant case, we first define some notation
by quickly reviewing the ordinary parity violating
case. In this case there are
two Weyl neutrino fields per generation. Denote these
by $\nu_L$ and $\nu_R$. The $\nu_L$ field is a member
of a $SU(2)_L$ doublet while $\nu_R$ is a gauge singlet.
The usual Higgs doublet can couple the $\nu_L$ and $\nu_R$
together and its vacuum expectation value will generate a
Dirac mass term. Also, since we assume that $\nu_R$ is electrically
neutral it can have a bare Majorana mass term coupling it
to itself.
Thus we have two mass terms:
$${\cal L}_{mass} = 2m\bar \nu_L \nu_R +
M \bar \nu_R (\nu_R)^c + H.c.. \eqno (17)$$
Note that $M$ is a bare mass term, and can take any value, while
$m$ is a mass term which is generated when the electroweak
gauge symmetry is broken. It is usually assumed that $M \gg m$.
The mass matrix has the form:
$${\cal L}_{mass} = (\bar \nu_L,  (\bar \nu_R)^c)
\left(\begin{array}{cc}
0&m\\
m&M
\end{array}\right)\left(
\begin{array}{c}
(\nu_L)^c\\
\nu_R
\end{array}\right) + H.c. . \eqno (18)$$
Diagonalising this mass matrix yields two Majorana mass eigenstates
with masses $m^2/M$ and $M$ (assuming that $M \gg m$).
If we denote the mass eigenstates by $\nu_{light}$ and $\nu_{heavy}$
then they can be written in terms of the weak eigenstates as
follows:
$$\nu_{light L}= \cos\phi \ \nu_L + \sin\phi \ (\nu_R)^c,$$
$$\nu_{heavy R} = - \sin\phi \ (\nu_L)^c + \cos\phi \ \nu_R,
\eqno (19)$$
where $\tan\phi = m/M$. Thus in the limit $M \gg m$,
we see that the light state is
essentially $\nu_L$ while the heavy state is essentially $\nu_R$.

The see-saw model is a simple extension of the standard model.
As in the case of
the standard model, it is straightforward to make it exactly
parity invariant. In this case, there are
four Weyl neutrino fields per generation:
$\nu_{L}, \nu_{R}$ and their mirror images $ N_{R}, N_{L}$.
Under the parity symmetry $\nu_{L,R} \leftrightarrow \gamma_0
N_{R,L}$. Note that since $N_R$ is the parity partner of $\nu_L$
it belongs to a $SU(2)_2$ doublet; while $N_L$ being the parity
partner of $\nu_R$ is a gauge singlet. (Recall that
the gauge group is defined in Eq.(3) and the parity transformations
are given in Eq.(5)). Assuming the minimal Higgs sector of one
ordinary Higgs doublet and its mirror image,
then the following mass terms are allowed (where for simplicity
we examine only one generation):
$${\cal L}_{mass} = 2m_1(\bar \nu_L \nu_R + \bar N_R N_L) +
2m_2 (\bar \nu_L (N_L)^c + \bar N_R (\nu_R)^c)$$
$$+ M_1 (\bar \nu_R (\nu_R)^c + \bar N_L (N_L)^c)
+ M_2 (\bar \nu_R N_L + \bar N_L \nu_R) + H.c..
\eqno (20)$$
Note that $m_{1,2}$ are mass terms which arise from
spontaneous symmetry breaking, while $M_{1,2}$ are bare mass terms.
As in the case discussed above, we will assume that $M_{1,2} \gg
m_{1,2}$. We will also assume that the masses are real.
This is an important restriction and it is essentially
the only way in which we depart from the most general case.
{}From Eq.(20) we see that the mass matrix has the form:
$${\cal L}_{mass} = \bar \nu_L {\cal M} (\nu_L)^c + H.c.,
\eqno (21)$$
where
$$\nu_L = ((\nu_L)^c,  N_R,  \nu_R, (N_L)^c)^T, \eqno (22)$$
and
$${\cal M} = \left(\begin{array}{cccc}
0&0&m_1&m_2\\
0&0&m_2&m_1\\
m_1&m_2&M_1&M_2\\
m_2&m_1&M_2&M_1
\end{array}\right). \eqno (23)$$
The mass matrix can be simplified by changing to the parity
diagonal basis $\nu^{\pm}_{L} = {\nu_L \pm (N_R)^c \over
\sqrt{2}}$, and $\nu^{\pm}_{R} = {\nu_R \pm (N_L)^c \over
\sqrt{2}}$.
In this basis the mass matrix has the form
$${\cal M} = \left(\begin{array}{cccc}
0&0&0&m_{+}\\
0&0&m_{-}&0\\
0&m_{-}&M_{-}&0\\
m_{+}&0&0&M_{+}
\end{array}\right), \eqno (24)$$
where $m_{\pm} = m_1 \pm m_2 $ and
$M_{\pm} = M_1 \pm M_2 $ .
The mass matrix can now be easily diagonalised because
it is essentially two copies of the $2 \times 2$
mass matrix Eq.(18). In the limit $M_{\pm} \gg m_{\pm},$ the
mass matrix Eq.(24) has eigenvalues:
$$m_{+}^2/M_{+}, m_{-}^2/M_{-}, M_{+}, M_{-},
\eqno (25)$$
and eigenvectors (which are the mass eigenstates):
$$\nu_1 = \nu^{+}_{L}, \  \nu_2 = \nu^{-}_L,\
\nu_3 = \nu^{+}_R, \   \nu_4 = \nu^{-}_R. \eqno (26)$$
Thus, in the one generation case, there is effectively only
two state mixing:
$$\nu_L = {\nu^{+}_L + \nu^{-}_L \over \sqrt{2}} $$
$$\  \     = {\nu_1 + \nu_2 \over \sqrt{2}}. \eqno (27)$$
Thus in this case the averaged neutrino oscillation probability
function ${\cal P}$
(defined in Eq.(13)) is $1/2$. In the physical case of three
generations, in general $\nu_L^{+}$ and $\nu_L^{-}$ will each
be linear combinations of three mass eigenstates:
$$\nu_{eL} = {\alpha_1 \nu_1^{'} + \alpha_2 \nu_2^{'} +
\alpha_3 \nu_3^{'}
\over \sqrt{2}} + {\beta_1 \nu_1^{''} + \beta_2 \nu_2^{''}+ \beta_3
\nu_3^{''} \over \sqrt{2}}, \eqno (28)$$
where $\alpha_1^2 + \alpha_2^2 + \alpha_3^2 = \beta_1^2 + \beta_2^2
+ \beta_3^2 = 1$ and $\nu_i^{'}, \nu_i^{''}$ are the six light
physical mass eigenstates.
In this case, the effect of
the neutrino oscillations averaged over many oscillations is to
reduce the electron neutrino flux by a factor $\cal P$
where
$$1/6 \le{\cal P} \le 1/2. \eqno (29)$$
To determine the precise value of ${\cal P}$ one must know
the actual values of the parameters $\alpha_i, \beta_i$ which can
only be determined if the neutrino mass matrix is known.
Note however that if intergenerational mixing
follows the pattern suggested by the quarks
(i.e. small intergenerational mixing angles) then one would
expect that the one generation result should be a good approximation
and that $\cal P$ should be nearly $1/2$.

Our second example is also a version of the see-saw model which
involves adding only three gauge singlets $N_{iR}$ (rather
than six as in the previous example above) which are parity
diagonal (i.e. under the parity transformation $N_{iR}
\leftrightarrow \gamma_0 (N_{iR})^c$).
In this case, for one generation, we have the following mass
terms (assuming the minimal Higgs sector Eq.(6)):
$${\cal L}_{mass} = m \left( \bar \nu_L N_R +  \bar \nu_R (N_R)^c
\right)
+ M \bar N_R (N_R)^c + H.c. \eqno (30)$$
where the parity transformation (Eq.(5)) interchanges
$\nu_L \leftrightarrow
\gamma_0 \nu_R$ and $N_R \leftrightarrow \gamma_0 (N_R)^c$
(note that, as before
$\nu_L$ ($\nu_R$) is an ordinary neutrino (mirror neutrino)
and is the
$SU(2)_1$ ($SU(2)_2$) partner of the electron (mirror electron).)
In this model, for the simplified case of one generation,
the phases of the fields can be chosen so that all of the
masses are real. Thus the reality assumption of the neutrino
mass matrix is redundant unless there is intergenerational
mixing. If intergenerational mixing is small, which is possible
(c.f case of quarks), then the simplified case of one generation
will be a good approximation to the physical case of three
generations.
If we change to the parity diagonal basis, by defining
$\nu_{\pm} = {\nu_L \pm (\nu_R)^c \over \sqrt{2}}$, then
the mass matrix has the form:
$${\cal L}_{mass} = (\bar \nu_{-L}, \bar \nu_{+L},
(\bar N_R)^c)
\left(\begin{array}{ccc}
0&0&0\\
0&0&m\\
0&m&M\end{array}\right)
\left(\begin{array}{c}
(\nu_{-L})^c\\
(\nu_{+L})^c\\
N_R
\end{array}\right) + H.c. .
\eqno (31)$$
Assuming that $M \gg m$, then there are 3 Majorana mass
eigenstates $N_R$, $\nu_{+L}$ and $\nu_{-L}$ with eigenvalues
$M, m^2/M, 0$. In this one generation case, there is effectively
two-state mixing, $\nu_{eL} = \nu_{+L} + \nu_{-L}/\sqrt{2}$
and ${\cal P} = 1/2$.
For the physical case of three generations, in general $1/6 \le
{\cal P} \le 1/2$ (assuming that $M_i \gg m_j$), however as before,
if intergenerational mixing is suppressed then we would expect
${\cal P}$ to be close to 1/2.

\vskip .7cm
\noindent
{\large \bf V Atmospheric neutrino anomaly and concluding remarks}
\vskip .6cm
It may be possible to explain the atmospheric neutrino
anomaly (Eq.(2)) if the muon
neutrino oscillates into the mirror neutrino on its way down from
the atmosphere. The oscillation
length for the electron neutrino can be large enough so that it
does not have time to oscillate (but short enough so that the
solar neutrinos oscillate!). Obviously, since the atmosphere is
not very deep there will not be a large range of parameters in which
the muon neutrino will have time to oscillate (this is unlike
the case for solar neutrino oscillations).
Another problem with this solution is
that it does not seem to be consistent with upcoming muon events
coming from muon neutrinos travelling through the Earth.
However this conclusion relies
heavily on the absolute neutrino flux estimate [6, 13].
It may be possible that the absolute muon neutrino flux has been
over estimated, and that the atmospheric neutrino anomaly is an
indication that the muon neutrinos mix
nearly maximally with a mirror neutrino (which
is effectively sterile with regard to ordinary matter).
Of course this possibility is very speculative at the moment,
and we await further neutrino experiments with interest to see
whether the atmospheric neutrino anomaly is stable (i.e.
time independent).

To summarise, we have examined neutrino oscillations in the exact
parity model. We have shown that the effect of vacuum
neutrino oscillations can be large when neutrinos mix with their
mirror partners. We proved some general
results which are nearly independent of the details of the
neutrino mass matrix.
In particular we examined the possibility of massive neutrinos and
we showed (in the context of exact parity symmetry)
that if the neutrino mass matrix is real, then the neutrino flux
of a source of weak eigenstate neutrinos (such as the sun)
is reduced by a factor greater than 2 provided the neutrino
oscillation lengths are small enough (i.e. neutrino mass differences
are big enough) compared with the size of the source (the sun) or
detector (the Earth). The sun is very big, so there
is quite a large range of parameters (about 8 orders of
magnitude) in which the flux of electron neutrinos
as measured in the experiments would be reduced by a factor
of 2 or more. Thus, we argue that if the
solar neutrino deficit and/or atmospheric
neutrino anomalies are signals of some new physics,
then this new physics might be vacuum neutrino oscillations.
They can be naturally large if parity is a exact symmetry of nature.

\vskip 1cm
\noindent
{\bf \large Acknowledgements}
\vskip .5cm
\noindent
I would like to thank Henry Lew and Ray Volkas for many
discussions concerning exact parity.

\vskip 1.5cm
\noindent
{\large \bf References}
\vskip .7cm
\noindent
[1] R. Davis, in Proceedings of the 21st International cosmic
ray conference, ed. R. J. Protheroe (University of
Adelaide Press, Australia, 1990) Vol.2 p 143;
Kamiokande collabroration, K. S. Hirata et al, Phys. Rev.
Lett. 65, 1297 (1990); Phys. Rev. D44, 2241 (1991);
GALLEX Collaboration, P. Auselmann et al; Phys. Lett.
B285, 376 (1992); GALLEX Collaboration, presented by F. van
Feilitzsch, in the Proceedings of the International conference
``Neutral currents twenty years later", Paris 1993, to appear;
SAGE Collaboration, A. I. Abazov et al, Phys. Rev. Lett. 67, 3332
(1991).

\vskip .5cm
\noindent
[2] J. N. Bachall et al, Rev. Mod. Phys. 54, 767 (1982);
J. N. Bahcall and R. K. Ulrich, Rev. Mod. Phys. 69, 297 (1988);
J. N. Bahcall and M. H. Pinsonneault, Rev. Mod. Phys. 64, 885 (1992).

\vskip .5cm
\noindent
[3] S. Turck-Chieze et al, Ap. J. 335, 415 (1988).

\vskip .5cm
\noindent
[4] D. R. O. Morrison, Int. J. of Mod. Phys. D1, 281 (1992);
CERN preprint/93-98 (1993).

\vskip .5cm
\noindent
[5] Kamiokande Collaboration, K. S. Hirata et al, Phys. Lett. B280,
146 (1992);
IMB Collaboration, D. Casper et al, Phys. Rev. Lett. 66, 2561 (1991);
R. Becher-Szendy et al., Phys. Rev. D46, 3720 (1992).

\vskip .5cm
\noindent
[6] T. K. Gaisser, T. Stanev and G. Barr, Phys. Rev. D38, 85 (1988);
G. Barr, T. K. Gaisser and T. Stanev, Phys. Rev. D39, 3532 (1989).

\vskip .5cm
\noindent
[7] B. Pontecorvo, Sov. JETP 26, 984 (1968);
V. Gribov and B. Pontecorvo, Phys. Lett. B28, 493 (1969).

\vskip .5cm
\noindent
[8] L. Wolfenstein, Phys. Rev. D17, 2369 (1978);
S. P. Mikheyev and A. Yu. Smirnov, Nuovo. Cimento 9C, 17 (1986).

\vskip .5cm
\noindent
[9] R. Foot, H. Lew and R. R. Volkas, Mod. Phys. Lett. A7, 2567
(1992).

\vskip .5cm
\noindent
[10] R. Foot, H. Lew and R. R. Volkas, Phys. Lett. B272, 67 (1991).
The idea that parity can be a symmetry of nature was discussed
earlier by T. D. Lee and C. N. Yang, Phys. Rev. 104, 254 (1956);
A. Salam, Nuovo Cimento, 5, 299 (1957);
V. Kobzarev, L. Okun and I. Pomeranchuk, Sov. J. Nucl. Phys.
3, 837 (1966); M. Pavsic, Int. J. Theor. Phys. 9, 229 (1974);
See also Ref [11].

\vskip .5cm
\noindent
[11] S. L. Glashow, Phys. Lett. B167, 35 (1986); E. D. Carlson
and S. L. Glashow, Phys. Lett. B193, 168 (1987); see also
B. Holdom, Phys. Lett. B166, 196 (1985).

\vskip .5cm
\noindent
[12] Particle data group, Phys. Rev. D (1992).

\vskip .5cm
\noindent
[13] M. Honda et al, Phys. Lett. B248, 193 (1990);
H. Lee and Y. S. Koh, Nuovo Cim. 105B, 883 (1990);
D. H. Perkins, Nucl. Phys. B399, 3 (1993).
\end{document}